\documentstyle[epsf,psfig,12pt]{article}
\renewcommand{\baselinestretch}{1.1}

\begin{document}
\def\be{\begin{eqnarray}}
\def\en{\end{eqnarray}}
\def\up{\uparrow}
\def\dw{\downarrow}
\def\non{\nonumber}
\def\la{\langle}
\def\ra{\rangle}
\def\nc{N_c^{\rm eff}}
\def\vp{\varepsilon}
\def\vma{{_{V-A}}}
\def\vpa{{_{V+A}}}
\def\m{\hat{m}}
\def\fp{{f_{\eta'}^{(\bar cc)}}}
\def\half{{{1\over 2}}}
\def\pr{{\sl Phys. Rev.}~}
\def\prl{{\sl Phys. Rev. Lett.}~}
\def\pl{{\sl Phys. Lett.}~}
\def\np{{\sl Nucl. Phys.}~}
\def\zp{{\sl Z. Phys.}~}
\def\lsim{ {\ \lower-1.2pt\vbox{\hbox{\rlap{$<$}\lower5pt\vbox{\hbox{$\sim$}
}}}\ } }

\font\el=cmbx10 scaled \magstep2
{\obeylines
\hfill IP-ASTP-04-97
\hfill NTU-TH-97-09
\hfill August, 1997}

\vskip 1.5 cm

\centerline{\large\bf Nonfactorizable Effects in Exclusive 
Charmless $B$ Decays}
\medskip
\bigskip
\medskip
\centerline{\bf Hai-Yang Cheng}
\medskip
\centerline{ Institute of Physics, Academia Sinica}
\centerline{Taipei, Taiwan 115, Republic of China}
\medskip
\centerline{\bf B. Tseng}
\medskip
\centerline{Department of Physics, National Taiwan University}
\centerline{Taipei, Taiwan 106, Republic of China}
\bigskip
\bigskip
\bigskip
\centerline{\bf Abstract}
\bigskip
{\small 
  Nonfactorizable effects in charmless $B\to PP,~VP$ decays can be lumped 
into the effective parameters $a_i$ that are linear combinations of Wilson
coefficients, or equivalently absorbed into the effective number of colors 
$\nc$. Naive factorization with $\nc=3$ fails to explain the CLEO data of
$B^\pm\to\omega K^\pm$, indicating the first evidence for the importance of
nonfactorizable contributions to the penguin amplitude. The decays $B^\pm
\to\omega\pi^\pm$ dominated by tree amplitudes are sensitive to the
interference between external and internal $W$-emission diagrams.
Destructive interference implied by $\nc=\infty$ leads to a prediction
of ${\cal B}(B^\pm\to\omega \pi^\pm)$ which is about $2\sigma$ too small
compared to experiment. Therefore, the CLEO data of 
$B^\pm\to\omega K^\pm,~\omega\pi^\pm$ suffice to rule out $\nc=3$ and 
strongly disfavor $\nc=\infty$ for rare charmless $B$ decays. Factorization
based on $\nc\approx 2$ can accommodate the data of $B^\pm\to\omega K^\pm$, 
but it predicts a slightly smaller branching ratio of $B^\pm\to\omega
\pi^\pm$ with ${\cal B}(B^\pm\to\omega\pi^\pm)/{\cal B}(B^\pm\to\omega K^\pm)
=0.6\,$. We briefly explain why the $1/N_c$ expansion is still applicable to 
the $B$ meson decay once the correct large-$N_c$ counting rule for the 
Wilson coefficient $c_2(m_b)$ is applied.

}

\pagebreak
 
{\bf 1.}~~To describe the hadronic weak decays of mesons, the mesonic
matrix elments are customarily evaluated under the factorization hypothesis
so that they are factorized into the product of two matrix elements of
single currents, governed by decay constants and form factors. In the
naive factorization approach, the relevant Wilson coefficient functions
for color-allowed external $W$-emission (or so-called ``class-I") and 
color-suppressed (class-II) internal $W$-emission
amplitudes are given by $a_1=c_1+c_2/N_c$, $a_2=c_2+c_1/N_c$, respectively, 
with $N_c$ the number of colors. Inspite of its tremendous simplicity, naive
factorization encounters two major difficulties. First, it never works for
the decay rate of class-II decay modes, though it usually operates for
class-I transition. For example, the predicted decay rate of the 
color-suppressed decay $D^0\to\bar 
K^0\pi^0$ in the naive approach is too small when compared with experiment 
(for a review, see \cite{Cheng89}).
Second, the hadronic matrix element under factorization is
renormalization scale $\mu$ independent as the vector or axial-vector
current is partially conserved. Consequently, the amplitude $c_i(\mu)
\la O\ra_{\rm fact}$ is not truly physical as the scale dependence of
Wilson coefficients does not get compensation from the matrix elements.
The first difficulty indicates that it is inevitable and 
mandatory to take into account nonfactorizable contributions, especially 
for class-II decays, to render the color suppression of internal $W$
emission ineffective. In principle, the second difficulty also should
not occur since the matrix elements of four-quark 
operators ought to be evaluated in the same renormalization scheme as that for 
Wilson coefficients and renormalized at the same scale $\mu$. 

   Because there is only one single form factor (or Lorentz scalar) 
involved in the
class-I or class II decay amplitude of $B\,(D)\to PP,~PV$ decays ($P$: 
pseudoscalar meson, $V$:
vector meson), the effects of nonfactorization can be lumped into the
effective parameters $a_1$ and $a_2$ \cite{Cheng94}:
\footnote{As pointed out in \cite{Kamal94}, the general amplitue of 
$B(D)\to VV$ decay consists of three independent Lorentz scalars, 
corresponding to $S$-, $P$- and $D$-wave amplitudes. Consequently, it
is in general not possible to define an effective $a_1$ or $a_2$ unless
nonfactorizable terms contribute in equal weight to all partial wave 
amplitudes.}
\be
a_1^{\rm eff}=c_1(\mu)+c_2(\mu)\left({1\over N_c}+\chi_1(\mu)\right),\quad 
a_2^{\rm eff}=c_2(\mu)+c_1(\mu)\left({1\over N_c}+\chi_2(\mu)\right),
\en
where nonfactorizable contributions are characterized by the parameters 
$\chi_1$ and $\chi_2$. Taking the decay $B^-\to D^0\pi^-$ as an example, 
we have \cite{Soares,Kamal96,Neubert}
\be
\chi_1(\mu) =\vp_8^{(BD,\pi)}(\mu)+{a_1\over c_2}\vp_1^{(BD,\pi)}(\mu),\quad
\chi_2(\mu) =\vp_8^{(B\pi,D)}(\mu)+{a_2\over c_1}\vp_1^{(B\pi,D)}(\mu),
\en
where
\be
\vp_1^{(BD,\pi)}(\mu) &=& {\la D^0\pi^-|(\bar du)_\vma(\bar cb)_\vma|B^-
\ra_{ nf}\over \la D^0\pi^-|(\bar du)_\vma(\bar cb)_\vma|B^-\ra_f}=
{\la D^0\pi^-|(\bar du)_\vma(\bar cb)_\vma|B^-
\ra\over \la \pi^-|(\bar du)_\vma|0\ra\la D^0|(\bar cb)_\vma|B^-\ra}-1,  
\non\\
\vp_8^{(BD,\pi)}(\mu) &=& {1\over 2}\,{\la D^0\pi^-|(\bar d\lambda^a u)_\vma
(\bar c\lambda^a b)_\vma|B^-
\ra\over \la \pi^-|(\bar du)_\vma|0\ra\la D^0|(\bar cb)_\vma|B^-\ra},
\en
are nonfactorizable terms originated from color-singlet and color-octet
currents, respectively, $(\bar q_1q_2)_\vma\equiv \bar q_1\gamma_\mu(1-\gamma
_5)q_2$, and $(\bar q_1\lambda^a q_2)_\vma\equiv \bar q_1\lambda^a \gamma_\mu
(1-\gamma_5)q_2$. The subscript `f' and `nf' in Eq.~(3) stand for
factorizable and nonfactorizable contributions, respectively, and the
superscript $(BD,\pi)$ in Eq.~(2) means that the pion is factored out in the
factorizable amplitude of $B\to D\pi$ and likewise for the superscript
$(B\pi, D)$. In the large-$N_c$ limit, $\vp_1={\cal O}(1/N_c^2)$
and $\vp_8={\cal O}(1/N_c)$ \cite{Neubert}. Therefore, the 
nonfactorizable term $\chi$ in the $N_c\to \infty$ limit is dominated 
by color-octet current operators. Since $|c_1/c_2|\gg 1$, it is evident from 
Eq.~(1) that even a small amount of nonfactorizable contributions will have a
significant effect on the color-suppressed class-II amplitude.
Note that the effective 
parameters $a_i^{\rm eff}$ include all the contributions
to the matrix elements and hence are $\mu$ independent \cite{Neubert}.
If $\chi_{1,2}$ are universal (i.e. channel independent) in
charm or bottom decays, then we still have a new factorization scheme
in which the decay amplitude is expressed in terms of factorizable 
contributions multiplied by the universal effective parameters
$a_{1,2}^{\rm eff}$. The first
systematical study of nonleptonic weak decays of heavy mesons within 
the framework of the improved factorization was carried 
out by Bauer, Stech, and Wirbel \cite{BSW}. Phenomenological analyses
of two-body decay data of $D$ and $B$ mesons indicate that while
the generalized factorization hypothesis in general works reasonably well, 
the effective parameters $a_{1,2}^{\rm eff}$ do show some variation from
channel to channel, especially for the weak decays of charmed mesons
 \cite{Cheng94,Kamal96,Cheng96}.
An eminent feature emerged from the data analysis is that $a_2^{\rm eff}$ 
is negative in charm decay, whereas it becomes positive in bottom decay 
\cite{Cheng94,CT95,Neubert}:
\be
a_2^{\rm eff}(D\to\bar K\pi)\sim -0.50\,, \quad a_2^{\rm eff}(B\to D\pi)\sim 
0.26\,.
\en
It should be stressed that since the magnitude of $a_{1,2}$ depends on the 
model results for form factors, the above values of $a_2$ should be 
considered as representative ones.
The sign of $a_2^{\rm eff}$ is fixed by the observed destructive
interference in $D^+\to\bar K^0\pi^+$ and constructive interference in
$B^-\to D^0\pi^-$. Eq.~(4) then leads to
\be
\chi_2(\mu\sim m_c;~D\to \bar K\pi)\sim -0.36\,, \quad \chi_2(\mu\sim m_b;~B
\to D\pi)\sim 0.11\,.
\en
In general the determination of $\chi_2$ is easier and more reliable 
than $\chi_1$. The observation $|\chi_2(B)|\ll|\chi_2(D)|$ is consistent
with the intuitive picture that soft gluon effects become stronger when the
final-state particles move slower, allowing more time for significant
final-state interactions after hadronization \cite{Cheng94}.

   Phenomenologically, it is often to treat the number of colors $N_c$ as
a free parameter and fit it to the data. Theoretically, this amounts to
defining an effective number of colors by 
\be
1/N_c^{\rm eff}\equiv (1/N_c)+\chi.
\en
It is clear from Eq.~(5) that
\be
N_c^{\rm eff}(D\to \bar K\pi) \gg 3,\quad N_c^{\rm eff}(B\to D\pi)
\sim 2.
\en
Consequently, the empirical rule of discarding 
subleading $1/N_c$ terms formulated in the large-$N_c$ approach \cite{Buras}
is justified for exclusive charm decay; the dynamical origin of the
$1/N_c$ expansion comes from the fact that the Fierz $1/N_c$ terms 
are largely compensated by nonfactorizable effects in charm decay.
Since the large-$N_c$ approach implies $a_2^{\rm eff}\sim c_2$ and since
$a_2^{\rm eff}$ is observed to be positive in $B^-\to D^{(*)}\pi^-(\rho
^-)$ decays, one may wonder why is the $1/N_c$ expansion no longer 
applicable to the $B$ meson ? Contrary to the common belief,
a careful study shows this is not the case. As pointed out in \cite{Neubert}, 
the large-$N_c$ color counting rule for the Wilson
coefficient $c_2(\mu)$ is different at $\mu\sim m_b$ and $\mu\sim m_c$ due
to the presence of the large logarithm at $\mu\sim m_c$. More specifically,
$c_2(m_b)={\cal O}(1/N_c)$ and $c_2(m_c)={\cal O}(1)$. Recalling that
$c_1={\cal O}(1)$, it follows that in the large-$N_c$ limit \cite{Neubert}:
\be
a_2^{\rm eff}=\cases{ c_2(m_c)+{\cal O}(1/N_c)   & for~the~$D$~meson,  \cr
c_2(m_b)+c_1(m_b)\left({1\over N_c}+\vp_8(m_b)\right)+{\cal O}(1/N_c^3) &
for~the~$B$~meson.   \cr}
\en
Therefore, {\it a priori} the $1/N_c$ expansion does not demand a 
negative $a_2^{\rm eff}$ 
for bottom decay ! and $N_c^{\rm eff}(B\to D\pi)\sim 2$ is not in conflict
with the large-$N_c$ approach ! It should be remarked that although 
$\chi_2$ is positive in two-body decays of the $B$ meson, some theoretical 
argument suggests that it may become negative for high multiplicity 
decay modes \cite{Neubert}.

    Thus far the nonfactorization effect is discussed at the purely
phenomenological level. It is thus important to have a theoretical 
estimate of $\chi_i$ even approximately. Unfortunately, all existing
theoretical calculations based on the QCD sum rule \cite{BS}, though confirm
the cancellation between the $1/N_c$ Fierz terms and nonfactorizable soft
gluon effects \cite{blok1}, tend to predict a negative $\chi$ in $\bar 
B^0\to D^+\pi^-,~D^0\pi^0$ and $B\to J/\psi K(K^*)$ decays. This tantalizing 
issue should be
clarified and resolved in the near future. It is interesting to remark that,
relying on a different approach, namely,
the three-scale PQCD factorization theorem, to tackle the nonfactorization 
effect, one of us and Li \cite{Li} are able to
explain the sign change of $\chi_2$ from bottom to charm decays.

   For $B$ meson decay, the effective parameters $a_{1,2}^{\rm eff}$ have
been determined so far only for $B\to D^{(*)}\pi(\rho)$ and $B\to J/\psi K
^{(*)}$ where nonfactorizable effects amount to having $N_c
^{\rm eff}\sim 2$. Recently, several exclusive charmless rare $B$ decay 
modes have been reported for the first time by CLEO \cite{CLEO} and many of 
them are dominated by the
penguin mechanism. It is thus important to know (i) does the constructive
interference of tree amplitudes persist in class-III charmless $B$ decay ?
(class-III transitions receive contributions from both external 
and internal $W$ emissions), and 
(ii) is $N_c^{\rm eff}\sim 2$ still applicable to the penguin amplitude ?
Whether $\nc\sim 2$ or $\nc\sim\infty$ for $B$ decay to two light mesons
is still under debate. For example, predictions for exclusive charmless
$B$ decay are presented in \cite{Kramer} for $\nc=\infty$. Recently, it was
argued in \cite{Ali} that the CLEO data of two-body charmless $B$ decays can
be accommodated by $0\leq 1/\nc\leq 0.5$ with $\nc=\infty$ being more 
preferred.
In this Letter we shall demonstrate that naive factorization (i.e.
$\nc=3$) is ruled out by the CLEO 
data of $B^\pm\to\omega K^\pm$ and $\nc=\infty$ is strongly disfavored by
the data of $B^\pm\to\omega\pi^\pm$. This implies the applicability of $\nc
\sim 2$ to the rare charmless $B$ decays.

\vskip 0.5mm
{\bf 2.}~~In this section we will consider the
decay modes dominated by penguin diagrams in order to study their $\nc$
dependence. It was pointed out in \cite{Deandrea} that the parameters
$a_2,~a_3$ and $a_5$ are strongly dependent on $\nc$ and the rates dominated
by these coefficients can have large variation. For example, the decay widths
of $B^-\to \omega K^{(*)-}$, $B^0\to\omega K^0,~\rho K^{*0}$, $B_s\to\eta
\omega,~\eta\phi,~\omega\phi,\cdots,$ etc. have strong $N_c$ dependence 
\cite{Deandrea}. We shall see that the branching ratio of $B^-\to\omega K^-$ 
has its lowest value near $\nc\sim 3-4$ and hence the naive factorization
with $\nc=3$ is ruled out by experiment.

   Before proceeding we briefly sketch the calculational framework.
The relevant effective $\Delta B=1$ weak Hamiltonian is
\be
{\cal H}_{\rm eff}(\Delta B=1) = {G_F\over\sqrt{2}}\Big[ V_{ub}V_{uq}^*(c_1
O_1^u+c_2O_2^u)+V_{cb}V_{cq}^*(c_1O_1^c+c_2O_2^c)
-V_{tb}V_{tq}^*\sum^{10}_{i=3}c_iO_i\Big]+{\rm h.c.},
\en
where $q=u,d,s$, and
\be
&& O_1^u= (\bar ub)_\vma(\bar qu)_\vma, \qquad\qquad\qquad\qquad~~
O_2^u = (\bar qb)_\vma(\bar uu)_\vma, \non \\
&& O_{3(5)}=(\bar qb)_\vma\sum_{q'}(\bar q'q')_{\vma(\vpa)}, \qquad  \qquad
O_{4(6)}=(\bar q_\alpha b_\beta)_\vma\sum_{q'}(\bar q'_\beta q'_\alpha)_{
\vma(\vpa)},   \\
&& O_{7(9)}={3\over 2}(\bar qb)_\vma\sum_{q'}e_{q'}(\bar q'q')_{\vpa(\vma)},
  \qquad O_{8(10)}={3\over 2}(\bar q_\alpha b_\beta)_\vma\sum_{q'}e_{q'}(\bar 
q'_\beta q'_\alpha)_{\vpa(\vma)},   \non
\en   
where $O_3$-$O_6$ are QCD penguin operators and $O_{7}$-$O_{10}$ originate
from electroweak penguin diagrams.
As noted in passing, in order to ensure the renormalization-scale and -scheme
independence for the physical amplitude, the matrix elements of 4-quark
operators have to be evaluated in the same renormalization scheme as
that for Wilson coefficients and renormalized at the same scale $\mu$. 
Before utilizing factorization, it is necessary to take into account QCD
and electroweak corrections to matrix elements:
\be
\la O_i(\mu)\ra=\left[\,{\rm I}+{\alpha_s(\mu)\over 4\pi}\m_s(\mu)+{\alpha
\over 4\pi}\m_e(\mu)\right]_{ij}\la O_j^{\rm tree}\ra,
\en
so that $c_i(\mu)\la O_i(\mu)\ra= \tilde{c}_i\la O_i^{\rm tree}\ra$, where
\be
\tilde{c}_i=\left[\,{\rm I}+{\alpha_s(\mu)\over 4\pi}\m_s(\mu)+{\alpha\over 4
\pi}\m_e(\mu)\right]_{ji}c_j(\mu).
\en
Then the factorization approximation is applied to the hadronic matrix
elements of the tree operator $O^{\rm tree}$. Perturbative QCD and 
electroweak corrections to the matrices $\hat m_s$ and $\hat m_e$ have been
calculated in \cite{Buras92,Flei,Kramer,Ali}. Using the next-to-leading order
$\Delta B=1$ Wilson 
coefficients obtained in the 't Hooft-Veltman scheme and the naive dimension 
regularization scheme at $\mu=4.4$ GeV, $\Lambda^{(5)}_{\overline{\rm 
MS}}=225$ 
MeV and $m_t=170$ GeV in Table 22 of \cite{Buras96}, we obtain the
renormalization-scheme and -scale independent Wilson coefficients 
$\tilde c_i$ at $k^2=m_b^2/2$: 
\footnote{The values of $\tilde c_i$ given in (13) are slightly different 
from that shown in \cite{CT97}. Here we have included QCD vertex 
corrections to
$\tilde c_1,\cdots,\tilde c_6$ (see \cite{Ali} for the expressions of
vertex corrections). It should be stressed that long-distance
nonfactorizable effects have not been included so far.}
\be
&& \tilde{c}_1=1.187, \qquad\qquad\qquad\qquad\qquad \tilde{c}_2=-0.312,   
\non \\
&& \tilde{c}_3=0.0236+i0.0048, \qquad\qquad\quad\,\tilde{c}_4=-0.0547-i0.0143, 
\non  \\
&& \tilde{c}_5=0.0164+i0.0048, \qquad\qquad\quad\,\tilde{c}_6=-0.0640-i0.0143, 
\non  \\
&& \tilde{c}_7=-(0.0757+i0.0558)\alpha, \qquad \quad \tilde{c}_8=0.057\,
\alpha, \non  \\
&& \tilde{c}_9=-(1.3648+i0.0558)\alpha, \qquad \quad \tilde{c}_{10}=
0.264\,\alpha.
\en

   We now apply the Hamiltonian (9) and factorization to the decay $B^-\to
\omega K^-$ and obtain
\be
A(B^-\to\omega K^-) &=& {G_F\over\sqrt{2}}\Bigg\{ V_{ub}V_{us}^*\left(a_1
X_1+a_2X_{2u}+a_1X_3\right)   \non \\
&-& V_{tb}V_{ts}^*\Bigg[ \left(a_4+a_{10}-2(a_6+a_8){m_K^2\over (m_s+m_u)(m_b
+m_u)}\right)X_1   \non \\
&+& {1\over 2}(4a_3+4a_5+a_7+a_9)X_{2u}   \non \\
&+& \left(a_4+a_{10}-2(a_6+a_8){m_B^2\over (m_s+m_u)(m_b+m_u)}\right)X_3
\Bigg]\Bigg\},
\en
where $a_{2i}\equiv\tilde{c}_{2i}+{1\over N_c}\tilde{c}_{2i-1},~a_{2i-1}\equiv
\tilde{c}_{2i-1}+{1\over N_c}\tilde{c}_{2i}$, and 
$X_i$ are factorizable terms:
\be
X_1 &\equiv& \la K^-|(\bar su)_\vma|0\ra\la\omega|(\bar ub)_\vma|B^-\ra=-i
\sqrt{2}f_K m_\omega A_0^{B\omega}(m_K^2)(\vp\cdot p_B),   \non \\
X_{2q} &\equiv& \la \omega|(\bar qq)_\vma|0\ra\la K^-|(\bar sb)_\vma|B^-\ra=
-i\sqrt{2}f_\omega m_\omega F_1^{BK}(m_{\omega}^2)(\vp\cdot p_B),   \non \\
X_3 &\equiv& \la\omega K^-|(\bar su)_\vma|0\ra\la 0|(\bar ub)_\vma|B^-\ra,
\en
with $\vp$ the polarization vector of the $\omega$ meson, and
$A_1$, $F_1$  the form factors defined in \cite{BSW85}. Just as in the case
of tree amplitudes, one can show that nonfactorizable effects in the penguin 
amplitudes of $B\to PP,~VP$ decays can be absorbed into the effective
penguin coefficients. This amounts to replacing $N_c$ in the penguin 
coefficients $a_i$ ($i=3,\cdots,10$) by $(N_c^{\rm eff})_i$. (It must be
emphasized that the factor of $N_c$ appearing in any place other than 
$a_i$ should {\it not} be replaced by $\nc$.) For simplicity, we will assume
$(N_c^{\rm eff})_1\approx (N_c^{\rm eff})_2\cdots\approx (N_c^{\rm eff})_{10}$
so that the subscript $i$ can be dropped.
For $\nc=3$, the QCD-penguin
Wilson coefficients are numerically given by Re\,$a_3=0.0054$, Re\,$a_4=
-0.0468$, Re\,$a_5=-0.0049$, and Re\,$a_6=-0.0585$. From Eq.~(14) we see 
that a large cancellation occurs in the QCD penguin amplitude due to the
large compensation between $a_3$ and $a_5$, $a_4$ and $a_6$. Since $|V_{tb}
V^*_{ts}|\gg |V_{ub}V^*_{us}|$, the decay rate of $B^\pm\to\omega K^\pm$
has its minimum around $\nc\sim 3-4$ (see Fig.~1), as noticed in
\cite{Deandrea,Kramer} and analyzed in detail in \cite{Ali}.

  Neglecting the $W$-annihilation contribution denoted by $X_3$, and 
using $f_K=160$ MeV, $f_\omega=195$ MeV for decay constants, $A_0^{B\omega}
(0)=0.28,~F_1^{B K}(0)=0.34$ \cite{BSW85}, and dipole $q^2$ dependence
for form factors $A_0$ and $F_1$ \cite{CCW}, $m_u=5$ MeV, $m_d=10$ MeV, 
$m_s=175$ MeV, $m_b=5$ GeV for quark masses, $\tau(B^\pm)=(1.66\pm 0.04)$ 
ps \cite{Richman} for the charged $B$ lifetime, and $A=0.804,~\lambda=0.22$,
$\eta=0.30,~\rho=\pm 0.30$ \cite{CT97} for Wolfenstein parameters 
\cite{Wolf}, we obtain
the averaged branching ratios of $B^\pm\to\omega K^\pm$ defined by
\be
{\cal B}(B^\pm\to\omega K^\pm)\equiv {1\over 2}\left[{\cal B}(B^+\to\omega 
K^+)+{\cal B}(B^-\to\omega K^-)\right],
\en
in Table I and Fig.~1. We see that the prediction ${\cal B}(B^\pm\to\omega
K^\pm)=1.44\times 10^{-6}$ at $\nc=3$ and $\rho=-0.30$ is off by $2\sigma$
from the experimental result \cite{CLEO}
\be
{\cal B}(B^\pm\to\omega K^\pm)=\left(1.2^{+0.7}_{-0.5}\pm 0.2\right)
\times 10^{-5}.
\en
This shows that naive factorization with $\nc=3$ (or $\chi=0$) fails 
to explain the decay rate of
$B^\pm\to\omega K^\pm$. It is clear from Table I or Fig.~1
that, for $\rho<0$, $\nc=2$ is slightly better than $\nc=\infty$ and yields a
branching ratio of order $8.3\times 10^{-6}$, in agreement with experiment.
\footnote{Our conclusions for $\nc=2$ and $\nc=\infty$ are different from
that in \cite{Ali} which claimed that a value of $1/\nc$ in the range
$0.15\leq 1/\nc\leq 0.55$ is disfavored by the data of $B^\pm\to\omega K^\pm$
and that $0\leq 1/\nc\leq 0.15$ is preferred.}
Note that the electroweak penguin contribution is constructive at $\nc=2$
and destructive at $\nc=\infty$. For $\rho>0$, $\nc=\infty$ works better 
than $\nc=2$ and can accommodate the data. However, we have 
shown in \cite{CT97} that a positive $\rho$ is quite disfavored by data
as it predicts ${\cal B}(B^\pm\to\eta\pi^\pm)\sim
1\times 10^{-5}$, which is marginally larger than the CLEO measured upper
limit: $0.8\times 10^{-5}$ \cite{CLEO}.

   A similar strong $\nc$ dependence also can be observed in $B^0_d\to\omega 
K^0$ decay. 
\footnote{As noted before, nonfactorizable effects in $B\to VV$ generally 
cannot be absorbed into the effective parameters $a_i$. Hence, we will not 
discuss $B\to\omega K^*,~\rho K^*$ decays.}
The expression of its factorizable amplitude is simpler than 
the charged $B$ meson:
\be
A(B^0_d\to\omega K^0) &=& {G_F\over\sqrt{2}}\Bigg\{ V_{ub}V_{us}^*a_2X_{2u}
-V_{tb}V_{ts}^*\Bigg[\,{1\over 2}(4a_3+4a_5+a_7+a_9)X_{2u}  \non \\
&+& \left(a_4-{1\over 2}a_{10}-(2a_6-a_8){m_K^2\over 
(m_s+m_d)(m_b+m_d)}\right)X_1      \Bigg]\Bigg\}.
\en
The averaged decay rate $\Gamma(\stackrel{_{_{(-)}}}{B}\!\!{^0}\to\omega 
\stackrel{_{_{(-)}}}{K}\!\!{^0})\equiv{1\over 
2}[\Gamma(B^0\to\omega K^0)+\Gamma(\bar B^0\to\omega\bar K^0)]$ is minimal
near $\nc\sim 4$ (see Fig.~2). The branching ratio predicted by $\nc=\infty$ 
is slightly larger than that by $\nc=2$ (see also Table I).

\vskip 0.4cm
\begin{table}
{{\small Table I. Averaged branching ratios for charmless $B$ decays, where 
``Tree" refers to branching ratios from 
tree diagrams only, ``Tree+QCD" from tree and QCD penguin diagrams, and 
``Tree+QCD+QED" from tree, QCD and electroweak (EW) penguin diagrams. 
Predictions are made for $k^2=m_b^2/2$, 
$\eta=0.30,~\rho=0.30$ (the first number in parentheses) and $\rho=-0.30$ 
(the second number in parentheses).}
{\footnotesize
\begin{center}
\begin{tabular}{|l|c|c c c |c|} \hline
Decay & $\nc$ & Tree & Tree$+$QCD & Tree$+$QCD$+$EW & Exp. \cite{CLEO}\\ 
\hline 
 & 2 & $6.53\times 10^{-7}$ & $(2.43,~6.82)\,10^{-6}$ &
$(3.28,~8.31)\,10^{-6}$ &  \\
$B^\pm\to\omega K^\pm$ & 3 & $4.49\times 10^{-7}$ & $(2.58,~9.63)\,10^{-7}$ 
& $(0.27,~1.44)\,10^{-6}$ & $(1.2^{+0.7}_{-0.5}\pm 0.2)\times 10^{-5}$ \\
 & $\infty$ & $1.56\times 10^{-7}$ & $(9.49,~6.44)\,10^{-6}$ & 
$(8.46,~5.61)\,10^{-6}$ &  \\
\hline
 & 2 & $5.25\times 10^{-8}$ & $(3.24,~4.46)\,10^{-6}$ &
$(3.78,~5.09)\,10^{-6}$ &  \\
$\stackrel{_{_{(-)}}}{B}\!{^0}\to\omega\stackrel{_{_{(-)}}}{K}\!{^0}$ & 3 & 
$4.64\times 10^{-9}$ & $(1.37,~2.10)\,10^{-7}$ & $(2.89,~3.96)\,10^{-7}$ &  \\
 & $\infty$ & $6.45\times 10^{-8}$ & $(6.28,~8.16)\,10^{-6}$ & 
$(5.18,~6.89)\,10^{-6}$ &  \\
\hline
 & 2 & $9.71\times 10^{-6}$ & $(1.15,~0.53)\,10^{-5}$ &
$(1.16,~0.52)\,10^{-5}$ &   \\
$B^\pm\to\omega \pi^\pm$ & 3 & $6.23\times 10^{-6}$ & $(7.07,~3.93)\,10^{
-6}$ & $(7.16,~3.76)\,10^{-6}$ & $(1.2^{+0.7}_{-0.5}\pm 0.2)\times 10^{-5}$ \\
 & $\infty$ & $1.57\times 10^{-6}$ & $(1.49,~1.93)\,10^{-6}$ 
& $(1.52,~1.78)\,10^{-6}$ &  \\
\hline
\end{tabular}
\end{center} } }
\end{table} 
\vskip 0.4cm

{\bf 3.}~~In the last section we have demonstrated that $\nc=3$ is ruled out 
as it fails to describe the decays $B^\pm\to\omega K^\pm$. This means 
that it is mandatory to take into account the nonfactorizable effect in 
the charmless $B$ decay amplitude. However, it is not easy to discern 
between $\nc=\infty$ and $\nc=2$ from $B\to\omega K$ decays, though the 
latter can explain
the data of $B^\pm\to\omega K^\pm$ and is more preferred. 
It is thus important to have a 
more decisive test on $\nc$. For this purpose, we shall focus in this 
section the decay modes dominated by the tree diagrams and sensitive to
the interference between external and internal $W$-emission amplitudes.
The fact that $\nc=2$ ($\nc=\infty$) implies constructive (destructive) 
interference will enable us to differentiate between them. Good
examples are the class-III modes: $B^\pm\to \pi^0\pi^\pm,~\eta\pi^\pm,~\pi^0
\rho^\pm,~\omega\pi^\pm,\cdots$. Since
$B^-\to\omega\pi^-$ is the only channel in the list that has been measured by
CLEO, we will first investigate this mode.

Under factorization, the decay amplitude of $B^-\to\omega\pi^-$ is given by
\be
A(B^-\to\omega\pi^-) &=& {G_F\over\sqrt{2}}\Bigg\{ V_{ub}V_{ud}^*\left(a_1
X'_1+a_2X'_{2u}+a_1X'_3\right)  \non \\
&-& V_{tb}V_{td}^*\Bigg[ \left(a_4+a_{10}-2(a_6+a_8){m_\pi^2\over (m_b+m_u)
(m_u+m_d)}\right)X'_1   \non \\
&+& {1\over 2}\left(4a_3+2a_4+4a_5+a_7+a_9-a_{10}\right)X'_{2u}   \non \\
&+& \left(a_4+a_{10}-2(a_6+a_8){m_B^2\over (m_b+m_u)
(m_u+m_d)}\right)X'_3 \Bigg]\Bigg\},
\en
with the expressions of $X'_i$ similar to (15). Since
\be
V_{ub}V_{ud}^*=A\lambda^3(\rho-i\eta), \quad V_{cb}V_{cd}^*=-A\lambda^3,
\quad V_{tb}V_{td}^*=A\lambda^3(1-\rho+i\eta),
\en
in terms of the Wolfenstein parametrization \cite{Wolf}, are of the same 
order of magnitude, it is clear that $B^-\to \omega\pi^-$ is dominated by  
external and internal $W$ emissions and that penguin contributions are
suppressed by the smallness of penguin coefficients. In the limit of 
$\nc\to\infty$, we have $a_1=c_1$ and $a_2=c_2$, which in turn imply a
destructive interference of tree amplitudes in $B^\pm\to\omega\pi^\pm$. 
It is easily seen from Eq.~(13) that the interference becomes 
constructive when $\nc<3.8$\,. From Fig.~3 or Table I
we see that the averaged branching ratio of $B^\pm\to\omega\pi^\pm$ has its
lowest value of order $2\times 10^{-6}$ at $\nc=\infty$ and then increases 
with $1/\nc$. Since experimentally
\cite{CLEO}
\be
{\cal B}(B^\pm\to\omega\pi^\pm)=\left(1.2^{+0.7}_{-0.5}\pm 0.2\right)
\times 10^{-5},
\en
it is evident that $\nc=\infty$ is strongly disfavored by the data. Note that 
though the predicted branching ratio ${\cal B}(B^\pm\to\omega\pi^\pm)=
1.16\times 10^{-5}$ for $\nc=2$ and $\rho=0.30$ (see Table I) is in 
good agreement with experiment, we have discussed in 
passing that a positive $\rho$ seems to be ruled out \cite{CT97}.
For $\rho<0$, our prediction
${\cal B}(B^\pm\to\omega\pi^\pm)=0.52\times 10^{-5}$ is on the verge of
the lower side of the CLEO data. Of course, the
CLEO measurement can be more satisfactorily explained by having
a much smaller $\nc$, but this possibility is very unlikely as it 
implies a large nonfactorization effect in $B^\pm\to\omega\pi^\pm$. Recalling
that the magnitude of nonfactorizable term is $\chi\sim 0.1$ in $B\to D\pi$
decay and that the energy release in the process $B\to\omega\pi$ is  
larger than that in $B\to D\pi$, it is thus expected physically that $\chi
\lsim 0.1$ for the former. Using $\nc=2$, we find a slightly smaller branching
ratio for $B^\pm\to\omega\pi^\pm$ with
\be
{ {\cal B}(B^\pm\to\omega\pi^\pm)\over {\cal B}(B^\pm\to\omega K^\pm)}=0.61\,.
\en
Since theoretically it is difficult to see how the
branching ratio of $B^\pm\to\omega\pi^\pm$ can be enhanced from $0.5\times
10^{-5}$ to $1.2\times 10^{-5}$, it is thus important to have a refined
and improved measurement of this decay mode. 

 In analogue to the decays $B\to D^{(*)}\pi(\rho)$, the interference effect of
tree amplitudes in class-III charmless $B$ decay can be tested by measuring
the ratios:
\be
R_1\equiv 2\,{{\cal B}(B^-\to\pi^-\pi^0)\over {\cal B}(\bar B^0\to \pi^-\pi^+
)},\qquad R_2\equiv 2\,{{\cal B}(B^-\to\rho^-\pi^0)\over {\cal B}(\bar 
B^0\to \rho^-\pi^+)},\qquad R_3\equiv 2\,{{\cal B}(B^-\to\pi^-\rho^0)\over 
{\cal B}(\bar B^0\to \pi^-\rho^+)}.
\en
Since penguin contributions are very small, to a good approximation we have
\be
R_1 &=& {\tau(B^-)\over\tau(B^0_d)}\left(1+{a_2\over a_1}\right)^2,  \non\\
R_2 &=& {\tau(B^-)\over\tau(B^0_d)}\left(1+{f_\pi\over f_\rho}\,{A_0^{B\rho}
(m^2_\pi)\over F_1^{B\pi}(m^2_\rho)}\,{a_2\over a_1}\right)^2,  \non\\
R_3 &=& {\tau(B^-)\over\tau(B^0_d)}\left(1+{f_\rho\over f_\pi}\,{F_1^{B\pi}
(m^2_\rho)\over A_0^{B\rho}(m^2_\pi)}\,{a_2\over a_1}\right)^2.
\en
Evidently, the ratios $R_i$ are greater (less) than unity when the 
interference is constructive (destructive). Numerically we find
\be
R_1=\cases{1.74,  \cr 0.58, \cr} \quad R_2=\cases{1.40, \cr 0.80, \cr} \quad
R_3=\cases{2.50 & for~$\nc=2$,  \cr  0.26 & for~$\nc=\infty$, \cr}
\en
where use of $\tau(B^0_d)=(1.55\pm 0.04)$ ps \cite{Richman}, $f_\rho=216$ MeV,
$A_0^{B\rho}(0)=0.28$ \cite{CCW} has been made. Hence, a measurement of
$R_i$ (in particular $R_3$), which has the advantage of being independent of
the parameters $\rho$ and $\eta$, will constitute a very useful test on the
effective number of colors $\nc$.

   We would like to stress once again that the observation that $\nc=\infty$ 
is very likely to be ruled out in charmless $B$ decay does {\it not}
imply the inapplicability of the large-$N_c$ approach to the $B$ meson 
case. As explained before, the correct large-$N_c$ counting rule for the 
Wilson coefficient $c_2(m_b)$ is proportional to $1/N_c$. Consequently, 
a nontrivial $a_2^{\rm eff}$, given by $c_2(m_b)+{1\over\nc}c_1(m_b)$, 
starts at the order of $1/N_c$ and hence $\nc$ cannot go to infinity.

\vskip 0.5mm
{\bf 4.}~~To conclude, by absorbing the nonfactorizable effects into the
effective number of colors $\nc$, we have shown that $\nc=3$ is ruled out
by the CLEO data of $B^\pm\to\omega K^\pm$, implying the inapplicability
of naive factorization to charmless $B$ decays, and that $\nc=\infty$
is strongly disfavored by the experimental measurement of $B^\pm\to\omega
\pi^\pm$, indicating a constructive interference in class-III charmless
$B$ decays.

  Since the energy release in charmless two-body decays of the $B$ meson 
is generally 
slightly larger than that in $B\to D^{(*)}\pi,~D^{(*)}\rho$, it is natural
to expect that $\nc$ for the $B$ decay into two light mesons is close
to $\nc(B\to D\pi)\approx 2$. We have shown that $\nc\approx 2$ can 
accommodate the data of $B^\pm\to\omega K^\pm$, 
but it predicts a slightly smaller branching ratio of $B^\pm\to\omega
\pi^\pm$ with ${\cal B}(B^\pm\to\omega\pi^\pm)/{\cal B}(B^\pm\to\omega K^\pm)
=0.6\,$.
Thus far, the discussion of nonfactorization in rare $B$ decay is at the 
purely phenomenological level. It remains to be a challenge, especially in 
the framework of the QCD sum rule, to compute it theoretically.

\bigskip\medskip
\noindent ACKNOWLEDGMENT:~~This work was supported in part by the National 
Science Council of ROC under Contract Nos. NSC86-2112-M-001-020 and
NSC86-2112-M-001-010-Y.

\renewcommand{\baselinestretch}{1.1}
\newcommand{\bi}{\bibitem}
%

\newpage

\begin{figure}[ht]
\vspace{1cm}

    \centerline{\psfig{figure=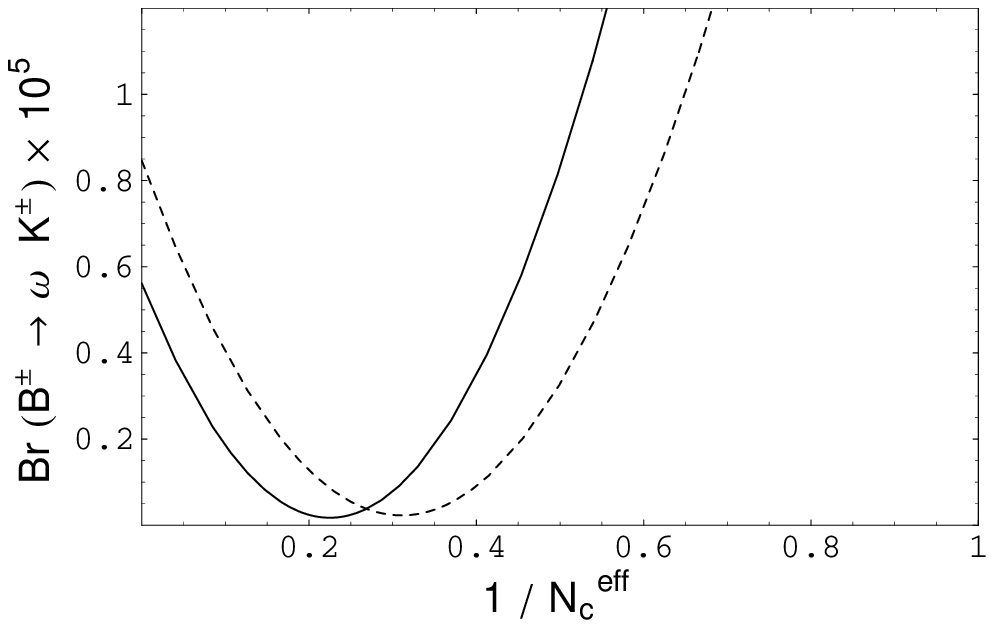,width=8cm,height=5.0cm}}
    \caption[]{\small The branching ratio of $B^\pm\to\omega K^\pm$ vs
       $1/\nc$. The solid and dashed curves are for $\rho=-0.30$ and
    $\rho=0.30$ respectively.}

\end{figure}

\begin{figure}[ht]
\vspace{1cm}

    \centerline{\psfig{figure=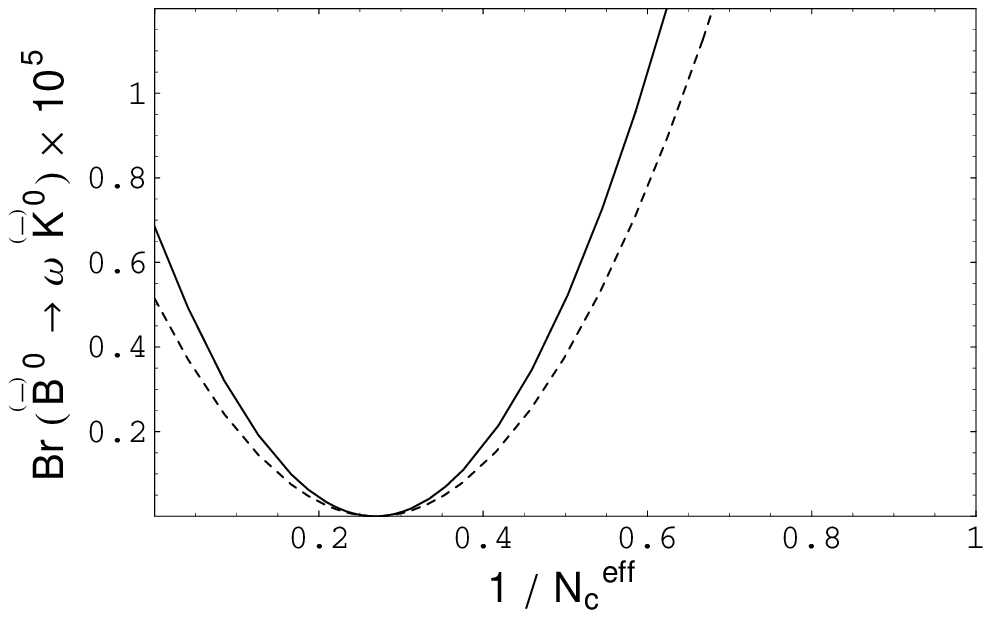,width=8cm,height=5.0cm}}
    \caption[]{\small Same as Fig.~1 except for $\stackrel{_{_{(-)}}}{B}\!\!
{^0}\to\omega \stackrel{_{_{(-)}}}{K}\!\!{^0}$.}

\end{figure}

\begin{figure}[ht]
\vspace{1cm}

    \centerline{\psfig{figure=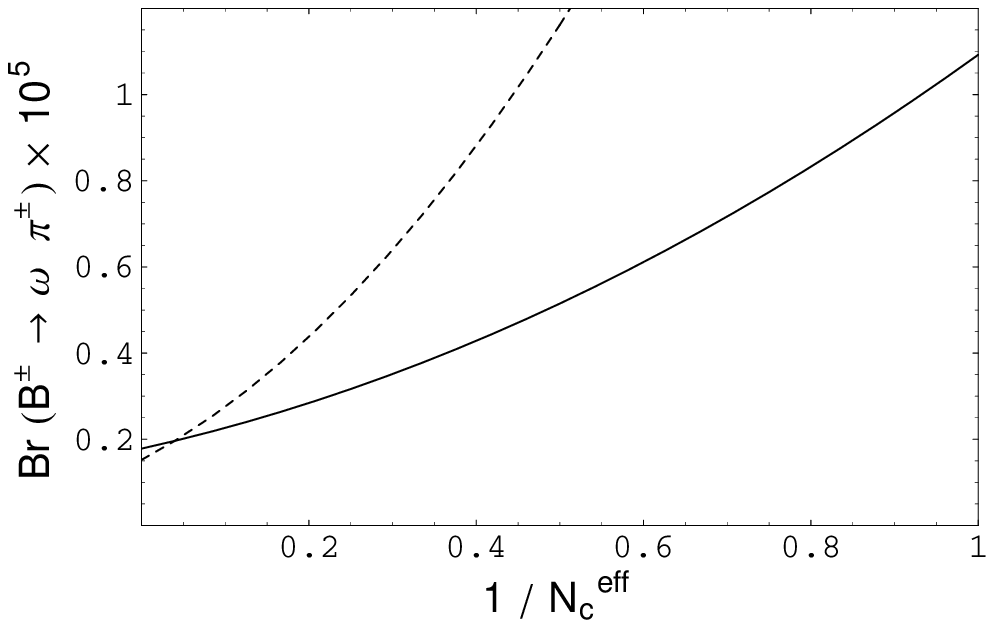,width=8cm,height=5.0cm}}
    \caption[]{\small Same as Fig.~1 except for $B^\pm\to\omega \pi^\pm$.}

\end{figure}


\begin{thebibliography}{99}
%

\bi{Cheng89} H.Y. Cheng, {\sl Int. J. Mod. Phys.} {\bf A4}, 495 (1989).

\bi{Cheng94} H.Y. Cheng, \pl {\bf B395}, 345 (1994); in {\it Particle Theory
and Phenomenology,}  XVII International Karimierz Meeting on Particle
Physics, Iowa State University, May 1995, edited by K.E. Lassila {\it et al}.
(World Scientific, Singapore, 1996), p.122~.

\bi{Kamal94} A.N. Kamal and A.B. Santra, Alberta Thy-31-94 (1994); \zp 
{\bf C72}, 91 (1996).

\bibitem{Soares}
J.M. Soares, \pr {\bf D51}, 3518 (1995);

\bi{Kamal96} A.N. Kamal, A.B. Santra, T. Uppal, and R.C. Verma, \pr 
{\bf D53}, 2506 (1996).

\bi{Neubert} M. Neubert and B. Stech, CERN-TH/97-99 [hep-ph/9705292], to 
appear in {\it Heavy Flavours}, edited by A.J. Buras and M. Lindner, 2nd ed. 
(World Scientific, Singapore).

\bibitem{BSW} M. Bauer, B. Stech, and M. Wirbel, \zp {\bf C34}, 103 (1987).

\bi{Cheng96} H.Y. Cheng, \zp {\bf C69}, 647 (1996).

\bi{CT95} H.Y. Cheng and B. Tseng, \pr {\bf D51}, 6295 (1995).

\bi{Buras}
A.J. Buras, J.-M. G\'erard, and R. R\"uckl, \np {\bf B268}, 16 (1986).

\bibitem{BS}
B. Blok and M. Shifman, \np {\bf B389}, 534 (1993);
A. Khodjamirian and R. R\"uckl, MPI-PhT/94-26 (1994); {\sl Nucl. Phys. (Proc.
Suppl.)} {\bf 39BC}, 396 (1995);
I. Halperin, \pl {\bf B349}, 548 (1995).

\bibitem{blok1}
B. Blok and M. Shifman, Sov. J. Nucl. Phys. {\bf 45}, 35, 301,
522 (1987).

\bi{Li} H.-n. Li and B. Tseng, NCKU-HEP-97-02 [hep-ph/9706441].

\bi{CLEO} CLEO Collaboration, talks presented by B. Behrens and J. Alexander
at {\it The Second 
International Conference on $B$ Physics and CP Violation}, March 24-27, 1997, 
Honolulu, Hawaii; F. W\"urthwein, CALT-68-2121 [hep-ex/9706010]; P.A.
Pomianowski, talk presented at {\it The Sixth Conference on the Intersections 
of Particle and Nuclear Physics}, Big Sky Montana, May 28, 1997.

\bi{Kramer} G. Kramer, W.F. Palmer, and H. Simma, \zp {\bf C66}, 429 (1995);
\np {\bf B428}, 77 (1994).

\bi{Ali} A. Ali and C. Greub, DESY 97-126 [hep-ph/9707251].

\bi{Deandrea} A. Deandrea, N. Di Bartolomeo, R. Gatto, F. Feruglio, and G.
Nardulli, \pl {\bf B320}, 170 (1994).

\bi{Buras92} A.J. Buras, M. Jamin, M.E. Lautenbacher, and P.H. Weisz, \np
{\bf B370}, 69 (1992); A.J. Buras, M. Jamin, and M.E. Lautenbacher, \np {\bf 
B408}, 209 (1993).

\bi{Flei} R. Fleischer, \zp {\bf C58}, 483 (1993).

\bi{Buras96} G. Buchalla, A.J. Buras, and M.E. Lautenbacher, {\sl Rev. Mod.
Phys.} {\bf 68}, 1125 (1996).

\bi{CT97} H.Y. Cheng and B. Tseng, IP-ASTP-03-97 [hep-ph/9707316].

\bi{BSW85} M. Wirbel, B. Stech, and M. Bauer, \zp {\bf C29}, 637 (1985).

\bi{CCW} H.Y. Cheng, C.Y. Cheung, and C.W. Hwang, \pr {\bf D55}, 1559 (1997).

\bi{Richman} J.D. Richman, hep-ex/9701014.

\bi{Wolf} L. Wolfenstein, \prl {\bf 13}, 562 (1984).


\end{thebibliography}
\end{document}